\documentclass[iop]{emulateapj}
\usepackage{graphicx}
\pdfoutput=1
\usepackage[english]{babel}
\usepackage{amsmath}
\usepackage{amssymb}
\usepackage{url}
\usepackage{natbib}
\usepackage{lineno}

\begin{document}

\shorttitle{PSR~J1023+0038 changes state}
\title{A state change in the missing link binary pulsar system PSR~J1023+0038}
\shortauthors{Stappers et~al.}

\author{B.~W. Stappers\altaffilmark{1},   A.~M. Archibald\altaffilmark{2}, J.~W.~T. Hessels\altaffilmark{2,3}, C.~G. Bassa\altaffilmark{2}, S. Bogdanov\altaffilmark{4},  G.~H. Janssen\altaffilmark{2}, V.~M. Kaspi\altaffilmark{5},  A.~G. Lyne\altaffilmark{1}, A. Patruno\altaffilmark{6,2}, S. Tendulkar\altaffilmark{7}, A.~B. Hill\altaffilmark{8,9}, T. Glanzman\altaffilmark{8}}
\altaffiltext{1}{Jodrell Bank Centre for Astrophysics, School of Physics and Astronomy, The University of Manchester, Manchester M13 9PL, UK}
\altaffiltext{2}{ASTRON, the Netherlands Institute for Radio Astronomy, Postbus 2, 7990 AA Dwingeloo, the Netherlands}
\altaffiltext{3}{Astronomical Institute A.Pannekoek, University of Amsterdam, 1098XH, Amsterdam, the Netherlands}
\altaffiltext{4}{Columbia Astrophysics Laboratory, Columbia University, 550 West 120th Street, New York, NY 10027, USA}
\altaffiltext{5}{McGill University, 3600 University Street, Montreal, QC H3A 2T8, Canada}
\altaffiltext{6}{Leiden Observatory, Leiden University, PO Box 9513, NL-2300 RA Leiden, the Netherlands}
\altaffiltext{7}{Space Radiation Laboratory, California Institute of Technology, 1200 E California Blvd, MC 249-17, Pasadena, CA 91125, USA}
\altaffiltext{8}{W. W. Hansen Experimental Physics Laboratory, KIPAC, Department of Physics and SLAC National
Accelerator Laboratory, Stanford University, Stanford, CA 94305, USA}
\altaffiltext{9}{Faculty of Physical \& Applied Sciences, University of Southampton, Highfield, Southampton, SO17 1BJ, UK}

\begin{abstract}

We present radio, X-ray, and $\gamma$-ray observations which reveal
that the binary millisecond pulsar / low-mass X-ray binary transition
system PSR~J1023+0038 has undergone a transformation in state.
Whereas until recently the system harbored a bright millisecond radio
pulsar, the radio pulsations at frequencies between 300 to 5000\,MHz have now become
undetectable. Concurrent with this radio disappearance, the $\gamma$-ray
flux of the system has quintupled.  We conclude that, though the radio
pulsar is currently not detectable, the pulsar mechanism is still
active and the pulsar wind, as well as a newly formed accretion disk,
are together providing the necessary conditions to create the 
$\gamma$-ray increase.  The system is the first example of a
transient, compact, low-mass $\gamma$-ray binary and will continue to
provide an exceptional test bed for better understanding the formation
of millisecond pulsars as well as accretion onto neutron stars in
general.

\end{abstract}



\section{Introduction}

The pulsar ``recycling'' scenario \citep{acrs82,rs81} proposes that
the rapid rotation rates of millisecond pulsars (MSPs\footnote{We use
  the term MSP for a neutron star rotating with a period $P_{\rm spin}
  \lesssim 30$\,ms and with an inferred surface magnetic field $B_{\rm
    surf} \lesssim 10^{10}$\,G.  Such sources derive their power from
  the star's rotational kinetic energy loss and necessarily produce
  observable radio pulsations.  In an increasing number of cases
  pulsed magnetospheric emission is also detected in keV X-rays and
  MeV$-$GeV $\gamma$-rays.})  originate from a period of mass
accretion from a binary companion in a low-mass X-ray binary
(LMXB). The idea that such systems could also result in the complete
ablation of the companion star and produce isolated MSPs --- like the
first MSP discovered, PSR B1937+21 \citep{bkh+82} --- seemed to be
confirmed by the discovery of the first eclipsing ``black widow''
pulsar PSR~B1957+20 \citep{fbb+90}. The eclipses of the radio
pulsations in this system showed that mass was being ablated from the
very low mass (M$_c\ll 0.1\,M_{\odot}$) companion \citep{rt91b}, however the mass loss rate appeared to be
insufficient to ever completely destroy the companion. The LMXB and
ablation phases were expected to be short-lived and so the subsequent
discovery of another black widow, PSR~J2051$-$0827 \citep{sbl+96},
seemed to challenge the connection between these systems and the
isolated MSPs.  Nonetheless, clear evidence that MSPs are indeed
spun-up in LMXBs came with the discovery of coherent X-ray pulsations
at a period of 2.5\,ms from the LMXB SAX J1808.4$-$3658 during an X-ray
outburst \citep{wv98}. To date there are some 15 of these accreting
X-ray millisecond pulsars (AMXPs) known \citep{pw12}.

With the rapid rate of MSP discoveries in the last few years
\citep[see his Fig. 1]{ran13}, the diversity of MSP binary companions has
continued to grow beyond the standard low-mass white dwarf or
ultra-low-mass black widow companions that were initially found.  This
variety has shown that the pulsar recycling process produces a rich
array of systems, some of which are exceedingly exotic
\citep[e.g.,][]{bbb+11}.  Furthermore, some of these new systems are
excellent case studies for better understanding the accretion-induced
spin-up of neutron stars.

Recently, a completely new class of eclipsing binary MSP systems,
dubbed ``redbacks'', with more massive (M$_\mathrm{c} \gtrsim
0.1$\,M$_{\odot}$) and likely non-degenerate companions have been
discovered (\citealt{rob11} and references therein). Targeted
searches for radio pulsars in unassociated \textit{Fermi}
$\gamma$-ray sources have been particularly fruitful, producing the
majority of these discoveries \citep{rap+12,hrm+11}.  Such searches
have also found many more examples of the previously known black widow
systems.  It has also been shown that during
the accretion stage the companion star can be ablated down to the mass
of a planet \citep{bbb+11,hnvj12} while at the same time (barely)
surviving complete destruction via ablation.  This again suggests that
in some cases the companion does not survive and an isolated MSP is
left behind.

With the wide variety of MSP systems now known, it is necessary to
better understand the various potential accretion and ablation
processes in LMXBs and eclipsing binary MSPs.  While most MSP systems
are clearly well past any accretion episode, the redbacks and black
widows appear to provide the closest available link to the LMXBs.
There are now two exceptional systems, the Galactic field binary
pulsar J1023+0038 (hereafter J1023; \citealt{asr+09}) and
PSR~J1824$-$2452I/IGR~J18245$-$2452 (hereafter M28I;
\citealt{pfb+13}), located in the globular cluster (GC) M28. Both are
providing us with a detailed look at the transition between these two
phases.  In fact, the behavior of these two systems in the last decade
has made it abundantly clear that the transition between an LMXB and
an MSP is not a sudden or unidirectional transformation.

J1023 was discovered in 2007 as a 1.7-ms radio pulsar in a 
4.8-hour circular-orbit eclipsing binary system. It was soon
recognized to be the same source as FIRST~J102347.6+003841, which had
originally been identified as a magnetic cataclysmic variable
\citep{bwb+02} but was subsequently shown to have been an LMXB with an
accretion disk during 2001 \citep{ta05}.  Later optical
and X-ray observations indicated that the source no longer possessed
an accretion disk \citep{wwp+04,hsc+06,asr+09,akb+10,bah+11}.  It was therefore
identified as the first object to have been seen to transition from an
LMXB to an MSP.  \cite{akh+13} also show that
since its discovery as a pulsar this ``original redback'' has
exhibited radio behavior typical of this class of eclipsing binary
MSPs. During this phase, X-rays are produced by the system
\citep{akb+10,bah+11} but they originate from a combination of pulsed
magnetospheric emission and an intra-binary shock between the
companion and MSP winds.

M28I links LMXBs and MSPs in a complementary way because it has been
seen to transition from a radio emitting MSP to an AMXP: i.e.\ it has shown accretion-powered
pulsations at the same rotational period as the previously known radio
pulsar \citep{pfb+13}. Moreover, it was suggested that this object has
been seen to swing between these states several times over the past
decade.

Here we report a sudden change of state in J1023, starting in 2013
June.  This change was heralded by the cessation of detectable pulsed
radio emission from the MSP and coincides with a dramatic, five-fold
increase in the $\gamma$-ray flux from the system.  Along with the
reported changes in the X-ray behavior \citep{pah+13} and the
emergence of double-peaked optical spectral lines \citep{hgs+13}, this
points to an accretion disk having re-formed in the system {\it but}
with a still-active pulsar mechanism also present.

\section{Observations and Data Reduction}

\subsection{Radio observations}

As part of a campaign to track the spin and orbital evolution of the
J1023 system, we are regularly observing it with the 76-m Lovell
Telescope (LT) at Jodrell Bank in the United Kingdom and the 94-m (equivalent) Westerbork
Synthesis Radio Telescope (WSRT) in the Netherlands.  We observed
using the LT on average once every 10 days, with a typical duration of
30\,min. Since 2013 July, we are observing with 1-hr integrations
approximately weekly.  The LT observations are centered at 1500\,MHz
and were recorded using both a digital filter bank (DFB) and, as of
2011 April, a coherent dedispersion system (ROACH) in parallel.  The
DFB and ROACH both provide 384\,MHz of usable bandwidth after
interference excision; the primary advantage of the ROACH data is
simply that it provides higher effective time resolution after
dedispersion using {\tt DSPSR} \citep{vb10}.  We used WSRT in
tied-array mode (where the signals from all the available dishes are
summed in phase) at central frequencies of 350 and 1380\,MHz with
respective bandwidths of 80 and 160\,MHz.  For both observing bands
the data were coherently dedispersed using the PuMaII backend
\citep{ksv08}.  Typically we observed for 25\,min, but since 2013 July
a number of longer observations, up to a full orbit, have been made.
For both the LT and WSRT, data inspection and post processing,
including dedispersion optimization and interference removal, is done
using the {\tt PSRCHIVE}\footnote{http://psrchive.sourceforge.net/}
package \citep{vdo12}.  The archived data products have 10-s time
resolution to search for short timescale changes in brightness.

In addition to our regular LT/WSRT monitoring, we have taken two long
observations using the Green Bank Telescope (GBT) on 2013 August 11 and the Arecibo
telescope (AO) on 2013 August 28, at central frequencies of 2\,GHz and 4.5\,GHz,
respectively.  At the GBT, we used the GUPPI pulsar backend to
coherently dedisperse an 800-MHz band.  The accumulated profiles were
written to disk every 2.64\,s.  With AO, we ran seven Mock
spectrometers, each recording 172\,MHz, in parallel which spanned
the total available 1-GHz band after removing overlap.  The band was divided into
256 channels, recorded every 32.768\,$\mu$s. {\tt DSPSR} was used
to incoherently dedisperse the data and fold it into 32 pulse phase
bins, writing out a profile every 100\,s. The various radio observing
systems are summarized in Table \ref{table:obs}.

\begin{table}[htdp]
\caption{Parameters of the radio telescopes used}
\begin{center}
\begin{tabular}{l|c|c|c|c|c}
\hline
Telescope & LT & WSRT & WSRT & GBT & AO \\
\hline \hline
Frequency (MHz) & 1500 & 350 & 1380  & 2300 & 4500 \\
Bandwidth (MHz) & 384 & 80 & 160   & 800 & 1000 \\
T$_{\rm sys}$ (K) & 27 & 150 & 25 & 25 & 25 \\
Gain (K Jy$^{-1}$) & 1.0 & 1.0 & 1.0 & 2.0 & 10.0 \\
\hline
\end{tabular}
\end{center}
\label{table:obs}
\end{table}%

\subsection{Gamma-ray observations}


J1023 is spatially associated with a $\gamma$-ray source detected while J1023 was active as a radio MSP (2FGL J1023.6+0040;  \citealt{thh+10,naa+12,akh+13}).  In light of this association,
and of the possibility of a change due to the disappearance of an
observable radio pulsar, we investigated the $\gamma$-ray light curve.
We retrieved the \textit{Fermi} Large Area Telescope \citep[LAT;][]{aaa+09c}  Pass 7 reprocessed photons with
energies of 100\,MeV $-$ 300\,GeV from the time range 2008 August 9 to 2013 November 18. We used the {\it Fermi} science tools version v9r32p5 for our analysis. We selected those photons with the {\tt SOURCE} event class, and
maximum zenith angle $100\degr$ to reduce contamination from atmospheric $\gamma$-rays.  The light curve analysis was then
performed using two approaches, which we now describe in turn.

The first approach follows the \textit{Fermi} Cicerone\footnote{http://fermi.gsfc.nasa.gov/ssc/data/analysis/documentation/Cicerone/} on
aperture photometry. We selected the ${>}1$\,GeV photons, for which the
point-spread function radius is ${\lesssim} 1$\degr \citep{aaa+12b}, that were within 1\degr\ of
J1023.  For reference, the nearest source in the 2FGL catalog
is 2\degr\ away.  We then applied a region of interest (ROI)
cut, filtered by the data selection {\tt (DATA\_QUAL==1) \&\& (LAT\_CONFIG==1) \&\& ABS(ROCK\_ANGLE)<52}, and computed good time intervals (GTIs). We binned the
resulting photons into 500-ks time bins and used the {\tt gtexposure}
tool with the {\tt P7REP\_SOURCE\_V15} instrument response functions, assuming a spectral index of $2.5$ \citep[the 2FGL catalog gives a spectral index for the associated source of $2.5\pm 0.3$;][]{naa+12}, to compute the
effective exposure within each bin. This left us with 260 photons before the disappearance and 63 after.
We then re-binned the light
curve and exposure values into 2.5-Ms bins and plotted the resulting
light curve, assuming Poisson errors.

The second approach follows the \textit{Fermi} Cicerone on
binned likelihood fitting. We again applied the same ROI cut,
data selection, and GTIs as above to all our photons. We constructed live time
cubes, exposure maps, and source maps for a $40\degr\times 40\degr$ square region. We extracted an 
XML\footnote{Extensible Markup Language} sky model
for all sources within a $35\degr$ circular region centered on J1023 from the 2FGL sky survey data\footnote{http://fermi.gsfc.nasa.gov/ssc/data/access/lat/2yr\_catalog/}. 
This model uses a log-parabola spectral model for J1023 and the public diffuse models {\tt gll\_iem\_v05} 
and {\tt iso\_source\-v05.txt}. In this sky model we kept the normalization free for all sources within 
$6\degr$ of J1023; all other parameters, including normalizations for sources further away, were held fixed.  We used
this model to extract light curves for all sources
within 6\degr\ of J1023. This process involved making a
series of 5-Ms sub-selections from our photon list. We applied the
above spectral fitting process --- generating live time cubes, exposure
maps, and source maps, then fitting our amplitude model --- to each
sub-selection. For those bins in which J1023 had Test Statistic less than 9 we used the python likelihood tools 
to compute 95\% upper limits. Finally we plotted the reported flux and errors or upper limits for
J1023 within each sub-selection, along with long-term
average fluxes and their errors for the pre-disappearance and post-disappearance spans.

In addition to building a light curve, we compiled a list of photons
possibly originating from J1023. To do this, we selected all photons within
$30\degr$ of J1023 and used the tool
{\tt gtsrcprob}, with the spectral model taken from the 2FGL catalog described above, to assign
each photon a probability of having come from J1023.  
Using simulations we determined that the low-probability photons play no role in 
modulation testing, so we discarded all photons with a probability less than 0.001.
We used these photons to search for orbital or rotational modulation in the
$\gamma$-ray flux from J1023.

\subsection{X-ray observations}

To monitor the X-ray behavior of J1023, we used the publicly
available data recorded with the \textit{Swift}/BAT Hard X-ray monitor \citep{khc+13} to generate a
$15-50$\,keV X-ray light curve from the period 2013 June 1 to November
2, excluding the period from August 23 to September 9, during which
Solar constraints prevented observations. The daily \textit{Swift}/BAT
sensitivity has an average value of
$2\times10^{35}\rm\,erg\,s^{-1}$. Variations in the daily exposure
times, however, can change the sensitivity between
$9\times10^{34}\rm\,erg\,s^{-1}$ and $10^{36}\rm\,erg\,s^{-1}$. The
coded exposure time (i.e., the exposure time rescaled by the
fractional coding) varies between ${\sim}100$ and ${\sim}20000$
seconds.

\begin{figure*}[htbp]
\begin{center}
\includegraphics[width=\linewidth]{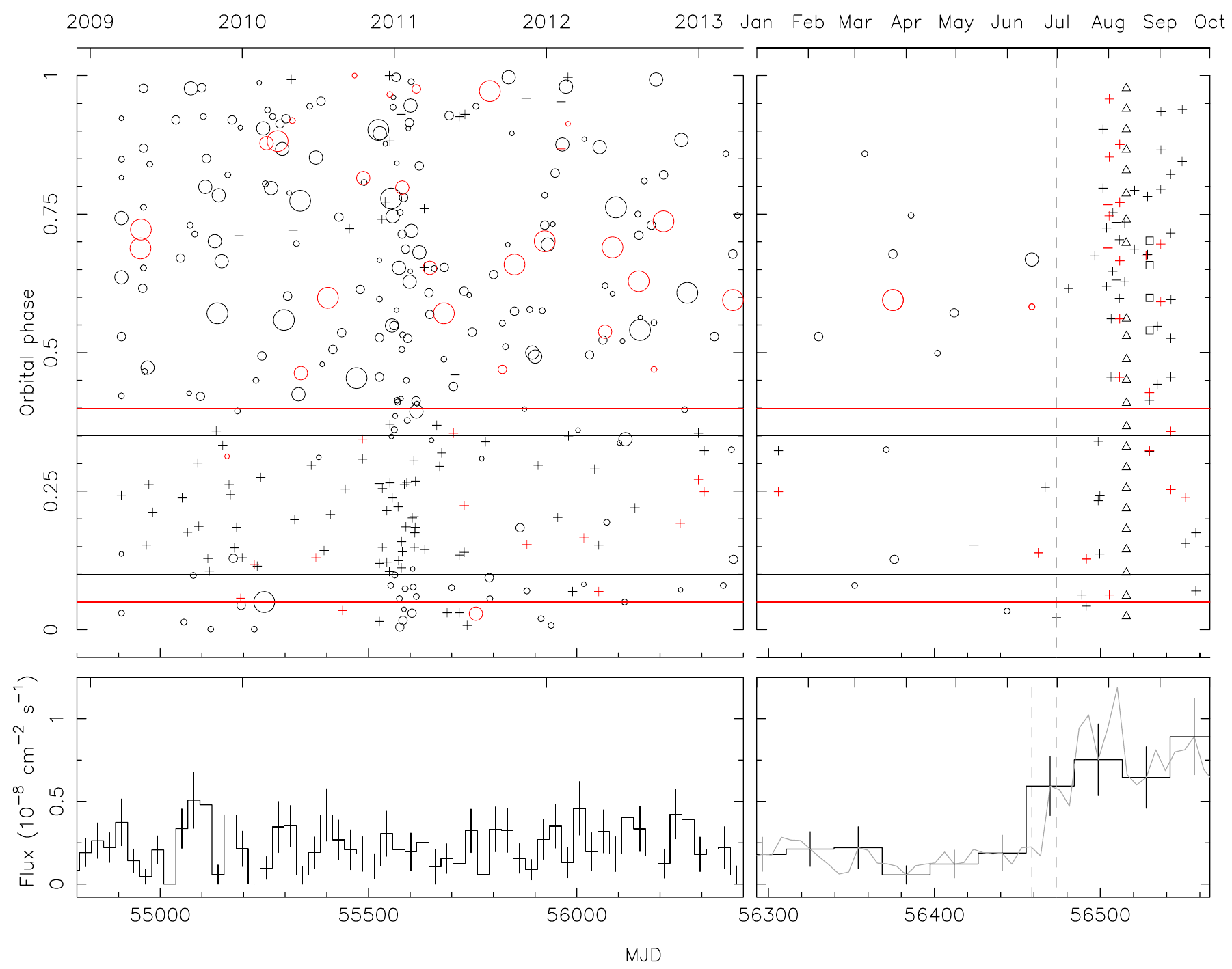}
\end{center}
\caption{Timeline for the state change in J1023. Top panel: 
  Radio observations of J1023 with the LT at 1500\,MHz and WSRT
  at 1380\,MHz (black symbols), WSRT at 350\,MHz (red symbols), GBT at
  2\,GHz (triangles) and Arecibo at 4.5\,GHz (squares). The
  observations are plotted against time and orbital phase, where an
  orbital phase of zero has the pulsar passing the ascending node. The
  left panel shows data from 2009 to 2013, while the right panel shows
  the data from 2013. Observations where the pulsar was not detected
  are denoted by plusses, triangles and squares, while detections are shown by circles, with
  the circle size indicating the signal-to-noise of the detection. The
  horizontal lines show the average eclipse duration at 1380 and
  1500\,MHz (black) and 350\,MHz (red). The vertical dashed lines indicate
  the last confirmed detection on 2013 June 15 of the pulsed signal
  and the first non-detection outside the known eclipse region on 2013 June 30.
  Bottom panel: 1--300 GeV $\gamma$-ray photon flux computed with aperture 
  photometry. The steps (solid line) show the flux averaged over 2.5-Ms 
  segments, with Poisson errors. The grey line shows the result of taking 
  the same 2.5-Ms averages with intermediate starting points, effectively 
  convolving the photon arrival time series with a 2.5-Ms top-hat function. }
\label{fig:radio}
\end{figure*}

\section{Results}

\subsection{Radio results}

A summary of the radio observations made with the LT and WSRT since
2009 is presented in Figure \ref{fig:radio}, where circles represent
clear detections of pulsed emission and and all other symbols indicate non-detections.  This plot
shows clearly the effect of the eclipses around orbital phase
0.25, when the pulsar is behind its companion.  The eclipse duration
approaches 50\% at low radio frequencies (red symbols); see \cite{akh+13} for a detailed discussion.  Variability away from
eclipse in the L-band (1380\,MHz and 1500\,MHz) observations (black
symbols) is caused by strong scintillation in the interstellar medium, shown
to have a bandwidth of about 100\,MHz \citep{akh+13} at this frequency, and
as expected for a pulsar with a dispersion measure of only 14.3\,pc\,cm$^{-3}$. The
350-MHz observations have a sufficiently large fractional bandwidth
that they are unaffected by scintillation, but the eclipses are longer
and there are also short duration, up to a few minutes long, eclipses at
other orbital phases. Apart from this variability, the most obvious
feature in Figure \ref{fig:radio} is that there has been no detection
of pulsed radio emission from the pulsar since mid June 2013. It may be
that J1023 is still an un-pulsed radio source but the observations reported here
are unable to determine if that is the case\footnote{While the WSRT is a synthesis telescope and
could in theory make an image of the source, it is an East-West array and the source is located at
the celestial equator which,  when combined with our relatively short integration times, makes imaging almost
impossible.}. 

Inspection of the regular LT timing observations in mid July revealed
that pulsations had not been detected since 2013 May 31 at
1500\,MHz. A check of the WSRT 350\,MHz and 1380\,MHz observations
showed that the source was last detected on 2013 June 15. Increasing
the cadence and duration of observations quickly revealed that the
pulsed radio emission was no longer seen at any of these
frequencies. Using the telescope and backend parameters given in Table
\ref{table:obs} we derive 10$\sigma$ flux density upper limits of 0.8
and 0.06\,mJy for WSRT 350-MHz and LT 1500-MHz observations,
respectively. These limits are more than an order of magnitude lower
than the typical flux densities of 16 and 0.9\,mJy measured between
2009 April and 2013 June. These limits rule out that the sudden
dissappearance is due to scintillation.

As shown in \cite{asr+09,akh+13}, the eclipse durations are
strongly frequency dependent (as well as varying somewhat between
orbits).  This motivated the higher-frequency GBT and AO observations
(the triangle and square symbols in Figure \ref{fig:radio},
respectively). A full orbit was sampled with the GBT observations and
the AO observations spanned orbital phases of 0.5$-$0.75, away from
the low-frequency eclipse region. No pulsed radio emission was
detected down to 10$\sigma$ flux density limits of 0.016 and
0.003\,mJy at 2\,GHz and 4.5\,GHz, respectively. These non-detections
confirm that the J1023 system changed to a different
state sometime between 2013 June 15 and 2013 June 30. The lack of observable radio
pulsations indicates that either  the radio pulsar mechanism has
been extinguished or  the radio pulsations are undetectable
because they are scattered in time or completely absorbed by an
increase of intra-binary material.  This state has persisted until at
least 2013 November.

\subsection{Gamma-ray results}

\begin{figure}[htbp]
\begin{center}
\includegraphics[width=\linewidth]{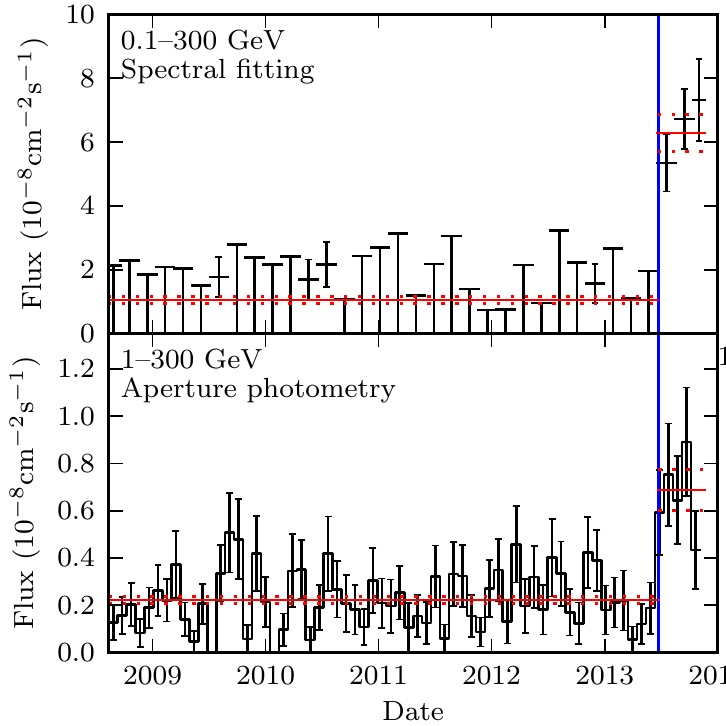}
\end{center}

\caption{$\gamma$-ray photon flux from J1023. Top panel: The
  100\,MeV--300\,GeV flux determined by spectral fitting, with
  uncertainties reported by the \textit{Fermi} tool {\tt gtlike}. 
  For time bins in 
  which the Test Statistic (${\sim}\sigma^2$) is less than 9 we instead 
  computed and plotted 95\% upper limits.  The
  blue line marks 2013 June 30, the approximate date of the disappearance. 
  The red lines show flux computed using all the before-disappearance 
  data and all the after-disappearance data, and the dotted lines show 
  the reported uncertainties on their values.
  Bottom panel: The 1--300\,GeV flux determined by aperture
  photometry, with Poisson uncertainties. The blue line marks the first
  radio non-detection as above. The red lines show the average flux
  before and after 2013 June 30, and the dotted lines show the
  (Poisson) uncertainties on their values.}

\label{fig:gamma}
\end{figure}

Figure \ref{fig:gamma} shows that the $\gamma$-ray flux has increased
dramatically since the radio disappearance.  This conclusion is
consistently reached using both of the analysis methods described in
\S 2.  Specifically, the average pre-disappearance $\gamma$-ray flux\footnote{Number(s) shown in parentheses represent the statistical uncertainty in the last digit(s) quoted. We do not expect systematic uncertainties to dominate and therefore do not consider them.}
obtained by spectral fitting is
$1.05(11)\times10^{-8}$\,cm$^{-2}$\,s$^{-1}$, while the average post-disappearance flux is 
$6.3(6)\times10^{-8}$\,cm$^{-2}$\,s$^{-1}$. By comparison, the fluxes obtained 
from aperture photometry are lower (because of the restriction to $1$--$300$ GeV 
photons and because not all ${>}1$ GeV photons fall within our aperture) and the background is higher (because the simple aperture cut allows 
more photons from other sources) but the average flux before the disappearance 
was  2.22(14)$\times10^{-9}$\,cm$^{-2}$\,s$^{-1}$, while the
post-disappearance average flux is 6.9(9)$\times
10^{-9}$\,cm$^{-2}$\,s$^{-1}$.

To better determine when the $\gamma$-ray flux started increasing, we
used our aperture-photometry data and combined 2.5-Ms chunks starting
every 500\,ks. The results are plotted in the lower panel of 
Figure \ref{fig:radio}. This
procedure unavoidably smoothes all features by 2.5\,Ms (about 1 month).  
Nonetheless, it
is clear that the $\gamma$-ray flux began rising at approximately
the same time that the radio pulsations disappeared. It is unclear,
however, whether the flux rose abruptly and then remained constant or
whether it rose gradually and is possibly continuing to rise.

To look for spectral changes in a nearly model-independent way, we
computed ``hardness ratios'' before and after the radio
disappearance. Examination of the aperture photometry photons revealed
that about half of them were above 1.6\,GeV, so we simply performed
aperture photometry as described above but with separate energy ranges
of 1$-$1.6\,GeV and 1.6$-$300\,GeV. A ``hardness ratio'' would then be
a ratio of (exposure-corrected) count rates in these two energy
ranges. The hardness ratio before the radio disappearance was
1.10(12), while after the disappearance it was 1.4(3). We do not claim 
any significant change in hardness.  Because of the low 
number of photons, we were not very
sensitive to spectral changes.  

We investigated orbital modulation of the $\gamma$-ray emission by
using {\tt tempo2} \citep{hem06} with the {\tt fermi} plugin to assign the
corresponding orbital phase to each probability-tagged photon.  We
used the weighted H test \citep{ker11} to look for periodic modulation,
obtaining a false positive probability of 0.66 --- that is, for purely
random phases there is a 66\% probability of having greater deviation
from uniformity than we detect. Testing with a simulated signal in
which all source photons are distributed according to a sine wave
produced a median false positive probability of 0.08. In other words,
even with this very strong modulation, we would have had only a 50\%
chance of obtaining a detection better than 1.4$\sigma$. So although we
have tested for orbital modulation, our non-detection rules out only
very sharply peaked orbital modulation. We investigated modulation at
the pulsar period using {\tt tempo2} to assign a pulse phase to
each photon, then testing for periodicity as above. The false positive
probability we obtained was 0.997. This method has the same lack of
sensitivity as our search for orbital pulsations, but it also has
another limitation: because our orbital ephemeris for J1023 was last
updated before its disappearance in radio, and because J1023 undergoes
apparently random orbital period variations \citep{akb+10,akh+13}, we should expect phase prediction errors as large as
${\sim}0.2$ turns to smear out any high harmonic content. It is
therefore difficult to rule out even very-sharply-peaked pulsations.

\subsection{X-ray results}

The \textit{Swift}/BAT finds only non-detections throughout the
monitoring, with all measurement points having a significance $< 3\sigma$. We simulated X-ray spectra with a power law of spectral
index between $\Gamma=1$ and $\Gamma=2$ and normalization set by using
the \textit{Swift}/XRT 0.5$-$10-keV luminosity reported in \citet{pah+13}.  For a spectral index
$\Gamma\ge1.3$ we infer a 15$-$50\,keV source luminosity below
$2\times10^{35}\rm\,erg\,s^{-1}$, which is compatible with the
\textit{Swift}/BAT non-detections and the currently measured spectral
index of the source ($\Gamma\simeq1.7$).

\section{Discussion}
We report that, some time between 2013 June 15 and June 30,  pulsations from
the millisecond pulsar in the J1023 binary system disappeared at
radio frequencies where they are otherwise easily
detectable. Concurrent with this disappearance we find that the
$\gamma$-ray flux of J1023 has increased five-fold. Subsequent
optical, UV and X-ray observations have shown that an accretion disk
has formed in the binary system \citep{hgs+13,pah+13}, signifying that
J1023 has undergone a transition from a binary MSP to an LMXB. The
present transition appears similar to that of 2000/2001; the archival
study by \citet{asr+09} shows the system transitioning from a radio
point source and G-type optical colors to having an accretion disk.

We are now fortunate to observe two redback systems that have
transitioned multiple times between LXMB and MSP states.  Although its
2013 outburst is a uniquely observed event for a known MSP, as a radio
pulsar M28I is fairly typical in a population of about 11 known
GC redbacks\footnote{http://www.naic.edu/$\sim$pfreire/GCpsr.html}.  This suggests that
other similar pulsars in the Galactic GC system will also undergo such
events in the coming years and decades.  These sources remain
challenging to study at optical and radio wavelengths, however,
because of the relatively large distance to the most massive GCs
($\sim 5-10$\,kpc).  Furthermore, since exchange interactions are
common in GCs, some GC redbacks host neutron stars that are likely
{\it not} orbited by their original binary companions.  Although they
are undoubtedly interesting test beds for studying the accretion
process, it is less clear whether they are useful for understanding
the long-term evolution from LMXBs to MSPs in the field.  In contrast,
the known field redbacks, of which J1023 was the first discovered, are typically
significantly less distant ($< 2$\,kpc) and offer better potential for
detailed multi-wavelength study.  There are now at least 7 redbacks
and 17 black widow systems known in the field, thanks in large part to
targeted searches of unassociated \textit{Fermi} sources
\citep{rap+12}.  Again, as a radio pulsar, J1023 is no longer very
atypical, with several similar examples, such as
PSR~J2215+5135 \citep{hrm+11}, which has nearly identical orbital
parameters and a likely similar, Roche-lobe-filling companion
\citep{bvr+13}.  Here as well, given proper radio, optical, and X-ray
monitoring, it is expected that it may not be long before these
similar systems show transitions like those we have observed from
J1023.  If, not, then the distinction of J1023 needs to be considered
more carefully.

Though J1023 and M28I show many striking similarities in their X-ray
behavior \citep{pah+13}, the former has yet to enter a truly energetic
outburst state in which the accretion flow is reaching the neutron
star surface.  Indeed, the 2013 June to October activity of J1023 is
arguably very similar to its past 2001 state (though no targeted X-ray
data was available at that time). An energetic outburst like that seen
in M28I \citep{pfb+13} can be ruled out for J1023 since 1996 because
it would have triggered an all-sky X-ray monitor.  M28I has also been
observed for $\sim200$\,ks by \textit{Chandra} in 2008 \citep{lbh+13},
switching several times between a bright ``active''
($4\times10^{33}\rm\,erg\,s^{-1}$) and a faint ``passive'' state
($6\times10^{32}\rm\,erg\,s^{-1}$). Therefore M28I can undergo rapid
X-ray flux variations similar to those currently seen in
J1023\footnote{Papitto et al.\ (2013) describes the 2008
  \textit{Chandra} active and passive states as an ``outburst''
  followed by a ``total quenching'' of the X-ray luminosity. We
  clarify, however, that the active/passive states cannot be
  considered as a standard outburst/quiescent cycle. The active state
  is indeed too faint and within the range of quiescent luminosities
  of LMXBs. Furthermore the passive state displays a significant X-ray
  luminosity well above the background level and therefore cannot be
  truly defined as quenching (see Linares et al.\ (2013) for an
  extended discussion.}.  If the X-ray flux variations observed in M28I
in quiescence (i.e., the 2008 \textit{Chandra} data, for which no
simultaneous radio observation is available) can be ascribed to the
presence of an accretion disk, then there is no specific reason to
presume that these two systems are fundamentally different.

\cite{shv70} proposed that an active radio pulsar in a contact binary
is able to prevent the formation of an accretion disk. However, M28I
has shown beyond any doubt a luminous X-ray outburst in 2013 April. In
all likelihood J1023 is also capable of undergoing a full accretion
episode, although the recurrence time could be on the order of
decades.  \cite{bpd+01} proposed that an LMXB can turn into a radio
pulsar either cyclically or indefinitely depending on the orbital
period of the system. According to Burderi's definition, J1023 falls
among the sources that cyclically alternate between an outburst and a
radio pulsar state. \cite{ea05} showed that an active pulsar does not
necessarily imply the complete destruction of the accretion disk if
the inner disk is truncated at 2--3 light cylinder radii.  It is
possible therefore that J1023 is an active pulsar surrounded by an
accretion disk truncated outside the light cylinder.  We suggest that
both J1023 and M28I can probably exhibit three states characterized by
i) an observable radio MSP and a low X-ray luminosity which is rotation powered ii) the presence
of an accretion disk in a quiescent-like LMXB system, and iii) an
AMXP, with high X-ray luminosity, powered by accretion, and potentially Type I X-ray bursts.

Regardless of the fact that the radio pulsar is currently not
detectable, it appears most likely that J1023's rotation-powered
pulsar mechanism, most importantly the relativistic particle wind that
is created, is still active. Before its recent state change, the
$\gamma$-rays from the system were quite likely of magnetospheric
origin, in analogy with other similar MSPs detected by \textit{Fermi}
LAT. As suggested by \citet{thh+10}, a portion of the $\gamma$-rays
could be from an intra-binary shock between the pulsar and its
companion star.  However, for the $\gamma$-ray data preceding 2013
June, \cite{akh+13} show that, while there is a marginally
significant detection of pulsations from J1023, there is no evidence for orbital modulation.

Searches for orbital-phase-dependent
$\gamma$-ray variability in other similar systems have also been
unsuccessful, again arguing that the observed emission from MSPs is
predominantly magnetospheric when the source is in the radio-loud
state.  Other similar $\gamma$-ray MSPs are stable in flux, as are
(almost) all young pulsars detected in $\gamma$-rays
\citep{aaa+13}. Recently the identification of $\gamma$-ray flux
variability from PSR~J2021+4026 \citep{abb+13} showed that some
$\gamma$-ray pulsars can undergo mode switches, reminiscent of recent
discoveries in the radio \citep{klo+06, lhk+10}.  More recently,
\cite{hhk+13} showed that for PSR~B0943+10 there is a correlated mode
switch in both radio and X-rays, providing strong evidence that a
sudden and global magnetospheric state change can occur. However,
while mode switching is a well known phenomenon observed in several
slowly rotating radio pulsars, there is as yet no evidence that
such magnetospheric switching occurs in MSPs.  Therefore, the
five-fold increase of the $\gamma$-ray flux in the current state is unlikely to be of magnetospheric origin or precipitated by a
change associated with the pulsar itself.

In light of this, it appears
more likely that the source of the increased $\gamma$-ray flux must be
an increased shock within the system.  The striking correlation with
the timing of the radio shut-off indicates that this rise in
$\gamma$-rays must be closely related to the reappearance of an
accretion disk in the system \citep{pah+13}, which is likely the
principal driving force behind the observed changes in the radio,
optical, X-ray, and $\gamma$-ray emission.  We suggest that to produce
the excess $\gamma$-rays requires both an active pulsar wind and an
accretion disk, but that there can be no active accretion onto the
neutron star, which would quench the pulsar mechanism and hence the
relativistic wind.

If the rotation-powered pulsar is still active, enshrouding of the
system by intra-binary material, causing severe scattering and/or
absorption, is the logical explanation for the absence of radio
pulsations. This would make J1023 a so-called ``hidden'' radio MSP, as
hypothesized by \citet{tav91}. In this scenario, the MSP is completely
enshrouded in the prodigious outflow from its close companion star,
rendering the pulsar perpetually eclipsed at radio frequencies.  At
the same time, the system generates strong un-pulsed shock emission in
X-rays and $\gamma$-rays, due to the interaction of the pulsar wind
with the outflow of matter \citep{tav93}.  Other recently discovered
eclipsing MSPs also exhibit compelling evidence for enshrouding
\citep{rs11,rrc+13}, though J1023 is the first instance where we see a
sudden increase in the high-energy emission, correlated with a radio
disappearance. Together, these systems support previous arguments that
a non-negligible fraction of MSPs may often be non-detectable at radio wavelengths
because of enshrouding. There is also some evidence that this becomes
more problematic for the fastest-spinning MSPs \citep{hrs+06,hrk+08}\footnote{Nonetheless, despite observational biases
  against finding fast-spinning MSPs as well as a possible populaton
  of fast-spinning, "hidden" MSPs, the spin frequency distribution of
  MSPs is clearly seen to drop off rapidly toward higher frequencies
   \citep[see his Fig 4.]{hrs+07}}.
  
Spin-down luminosity is a sensitive function of spin period ($\dot{\rm
  E} \propto \dot{\rm P}/\rm P^3$) and hence this may explain this
trend, all other factors, like orbital separation and companion type,
being equal. Indeed this could be one reason for the absence of a discovery of a sub-millisecond pulsar thus far, in spite of reasonable sensitivity to such sources in modern large-scale pulsar surveys, such as the PALFA survey \citep{cfl+06} and  the HTRU survey \citep{kjv+10} and the aforementioned LAT-directed searches.  On the other hand, the existence
of many such ''hidden"  fast radio MSPs is problematic in light of the absence
of detections of accreting MSPs with frequency higher than 619 Hz, in spite
of there being no obvious selection effects against finding them with X-ray
telescopes \citep{cmm+03}. J1023 is a 1.7-ms pulsar, the fifth fastest spinner known
in the Galactic field. Of the five fastest-spinning MSPs in the field,
three are known to eclipse and the other two are isolated.

With the re-appearance of an accretion disk, J1023 also resembles a
number of LMXB systems containing AMXPs.  It
has been suggested that some AMXPs reactivate as rotation-powered
pulsars in quiescence \citep{csm+98,cdc+04}. Apart from M28I, no radio
pulsations have been detected from any AMXP in quiescence, although
this could be a consequence of enshrouding as well. In addition, none
has yet been detected as $\gamma$-ray sources with \textit{Fermi}
LAT \citep{xw13}, although they are all significantly more distant
than J1023. It is thus plausible that quiescent AMXPs may also be
``hidden'' MSPs.

Given the nature of the enhanced $\gamma$-ray emission that has
appeared from J1023, one may consider what similarities exist with the
small population of $\gamma$-ray binaries, where a compact object is
orbited by a massive OB star \citep[e.g.][]{dub13}. Still the best understood
high-mass gamma-ray binary is PSR B1259$-$63, which contains a relatively rapidly-rotating (P=47 ms) young radio pulsar; its pulsar has $\dot{E}$
$\sim$20 times higher than that of PSR J1023+0038 yet has shown
$\gamma$-rays $\sim$200 times brighter, achieving near 100\% efficiency
in converting spin-down power to $\gamma$-rays near periastron, when the
pulsar's wind is shocked closest to the pulsar.  Though the orbital
compactness and companion mass in the case of J1023 is quite different
when compared with such systems, there may be similarities in some of
the physical mechanisms at play. This suggests PSR J1023+0038
could indeed continue to brighten.  It may also serve a similar prototype role
for a new class of low-mass $\gamma$-ray binaries. We note that given its increased $\gamma$-ray luminosity, its present $\gamma$-ray
efficiency, assuming an isotropic luminosity and taking $\dot{E} = 4.3 \times 10^{34}$~erg/s
\citep{dab+12}, is 0.14.  We therefore expect that the $\gamma$-ray luminosity
will rise at most another factor of 7, as beyond that it would exceed the available energy from pulsar spin-down.

Regardless of J1023's upcoming X-ray behavior, when it returns to an
MSP state then it will be possible to compare the radio pulsar timing
before and after the current active state. J1023 benefits greatly in
this regard from its relative brightness and the high cadence of our
recent radio monitoring observations. It also does not suffer from the
contaminating influence of acceleration in a GC gravitational
potential, as in the case of M28I.  After being spun-up to millisecond
periods, an accreting neutron star should enter a spin-down phase
where the mass transfer rate decreases due to the progressive
detachment of the donor radius from its Roche lobe. During this final
evolutionary phase the neutron star magnetosphere expands
substantially and a large fraction of the neutron star rotational
energy is lost via a propeller-like mechanism \citep{tau12}.  It is
possible therefore that what we are observing in J1023 is indeed this
very last phase of its life as an LMXB. In this case it should be
possible to observe a strong spin down during this active phase
i.e. well in excess of the pulsar magnetic dipole spin down. Even if
no accretion-powered X-ray pulsations are detected, this test can be
performed once the active phase is over by comparing the pre- and
post-active phase radio timing solutions.

In conclusion, we have demonstrated a new stage in the
back-and-forth transitioning that seems to characterize the redback
class of MSPs. For the first time, we have seen a bright radio MSP become undetectable while at the same time the system becomes
$\gamma$-ray bright. We argue that the pulsar mechanism remains
active in this system, but that the radio pulsations are obscured.
While higher-frequency radio observations have also been unsuccessful,
there is perhaps potential for detecting pulsed X-rays (not generated
by accretion) that will further confirm the continued activity of the
pulsar itself.  The detection or lack of accretion-powered X-ray
pulsations in forthcoming X-ray observations can easily disprove or
strengthen such an hypothesis. Similarly, planned
  radio interferometric observations to look for a continuum radio
  source are also important in this regard.

The analogy with the wider-orbit, much more massive companion
$\gamma$-ray binaries is tantalizing.  A multi-wavelength campaign
that follows J1023 back into its radio MSP state may better resolve the
state transition and associated radio/optical/X-ray/$\gamma$-ray
phenomenology.  Long-term, phase-coherent timing will also shed light
on the accretion torques experienced while the radio pulsar was
undetectable.

\acknowledgements{We thank H.A. Krimm for kind support in the use of
  the \textit{Swift}/BAT data. A.P. acknowledges support from the
  Netherlands Organization for Scientific Research (NWO) Vidi
  fellowship.  A.M.A. and J.W.T.H. acknowledge funding for this work
  from an NWO Vrije Competitie grant. The WSRT is operated by ASTRON
  with support from NWO. Pulsar observations with the Lovell Telescope are
  funded through a consolidated grant from STFC. The Fermi LAT is a pair conversion telescope designed to cover the
energy band from 20 MeV to greater than 300 GeV. It is the product of
an international collaboration between NASA and DOE in the U.S. and
many scientific institutions across France, Italy, Japan and Sweden. The GBT is operated by the National Radio Astronomy Observatory (NRAO).  NRAO is a facility of the NSF operated under cooperative agreement by Associated Universities, Inc. The Arecibo Observatory is operated
by SRI International under a cooperative agreement with the NSF (AST-1100968), and in alliance with
Ana G. M\'endez-Universidad Metropolitana, and the Universities Space Research Association. The $Fermi$ LAT Collaboration acknowledges support from a number of agencies and institutes for both development and the operation of the LAT as well as scientific data analysis. These include NASA and DOE in the United States, CEA/Irfu and IN2P3/CNRS in France, ASI and INFN in Italy, MEXT, KEK, and JAXA in Japan, and the K.~A.~Wallenberg Foundation, the Swedish Research Council and the National Space Board in Sweden. Additional support from INAF in Italy and CNES in France for science analysis during the operations phase is also gratefully acknowledged. ABH acknowledges that this research was supported by a Marie Curie International Outgoing Fellowship within the 7th European Community Framework Programme (FP7/2007--2013) under grant agreement no. 275861.}



\begin{thebibliography}{}
\expandafter\ifx\csname natexlab\endcsname\relax\def\natexlab#1{#1}\fi

\bibitem[{{Abdo} {et~al.}(2013){Abdo}, {Ajello}, {Allafort}, {Baldini},
  {Ballet}, {Barbiellini}, {Baring}, {Bastieri}, {Belfiore}, {Bellazzini}, \&
  et~al.}]{aaa+13}
{Abdo}, A.~A., {Ajello}, M., {Allafort}, A., {et~al.} 2013, \apjs, 208, 17

\bibitem[{{Ackermann} {et~al.}(2012){Ackermann}, {Ajello}, {Albert},
  {Allafort}, {Atwood}, {Axelsson}, {Baldini}, {Ballet}, {Barbiellini},
  {Bastieri}, {Bechtol}, {Bellazzini}, {Bissaldi}, {Blandford}, {Bloom},
  {Bogart}, {Bonamente}, {Borgland}, {Bottacini}, {Bouvier}, {Brandt},
  {Bregeon}, {Brigida}, {Bruel}, {Buehler}, {Burnett}, {Buson}, {Caliandro},
  {Cameron}, {Caraveo}, {Casandjian}, {Cavazzuti}, {Cecchi}, {{\c C}elik},
  {Charles}, {Chaves}, {Chekhtman}, {Cheung}, {Chiang}, {Ciprini}, {Claus},
  {Cohen-Tanugi}, {Conrad}, {Corbet}, {Cutini}, {D'Ammando}, {Davis}, {de
  Angelis}, {DeKlotz}, {de Palma}, {Dermer}, {Digel}, {Silva}, {Drell},
  {Drlica-Wagner}, {Dubois}, {Favuzzi}, {Fegan}, {Ferrara}, {Focke}, {Fortin},
  {Fukazawa}, {Funk}, {Fusco}, {Gargano}, {Gasparrini}, {Gehrels}, {Giebels},
  {Giglietto}, {Giordano}, {Giroletti}, {Glanzman}, {Godfrey}, {Grenier},
  {Grove}, {Guiriec}, {Hadasch}, {Hayashida}, {Hays}, {Horan}, {Hou}, {Hughes},
  {Jackson}, {Jogler}, {J{\'o}hannesson}, {Johnson}, {Johnson}, {Johnson},
  {Kamae}, {Katagiri}, {Kataoka}, {Kerr}, {Kn{\"o}dlseder}, {Kuss}, {Lande},
  {Larsson}, {Latronico}, {Lavalley}, {Lemoine-Goumard}, {Longo}, {Loparco},
  {Lott}, {Lovellette}, {Lubrano}, {Mazziotta}, {McConville}, {McEnery},
  {Mehault}, {Michelson}, {Mitthumsiri}, {Mizuno}, {Moiseev}, {Monte},
  {Monzani}, {Morselli}, {Moskalenko}, {Murgia}, {Naumann-Godo}, {Nemmen},
  {Nishino}, {Norris}, {Nuss}, {Ohno}, {Ohsugi}, {Okumura}, {Omodei},
  {Orienti}, {Orlando}, {Ormes}, {Paneque}, {Panetta}, {Perkins},
  {Pesce-Rollins}, {Pierbattista}, {Piron}, {Pivato}, {Porter}, {Racusin},
  {Rain{\`o}}, {Rando}, {Razzano}, {Razzaque}, {Reimer}, {Reimer}, {Reposeur},
  {Reyes}, {Ritz}, {Rochester}, {Romoli}, {Roth}, {Sadrozinski}, {Sanchez},
  {Saz Parkinson}, {Sbarra}, {Scargle}, {Sgr{\`o}}, {Siegal-Gaskins},
  {Siskind}, {Spandre}, {Spinelli}, {Stephens}, {Suson}, {Tajima}, {Takahashi},
  {Tanaka}, {Thayer}, {Thayer}, {Thompson}, {Tibaldo}, {Tinivella}, {Tosti},
  {Troja}, {Usher}, {Vandenbroucke}, {Van Klaveren}, {Vasileiou}, {Vianello},
  {Vitale}, {Waite}, {Wallace}, {Winer}, {Wood}, {Wood}, {Wood}, {Yang}, \&
  {Zimmer}}]{aaa+12b}
{Ackermann}, M., {Ajello}, M., {Albert}, A., {et~al.} 2012, \apjs, 203, 4

\bibitem[{{Allafort} {et~al.}(2013){Allafort}, {Baldini}, {Ballet},
  {Barbiellini}, {Baring}, {Bastieri}, {Bellazzini}, {Bonamente}, {Bottacini},
  {Brandt}, {Bregeon}, {Bruel}, {Buehler}, {Buson}, {Caliandro}, {Cameron},
  {Caraveo}, {Cecchi}, {Chaves}, {Chekhtman}, {Chiang}, {Chiaro}, {Ciprini},
  {Claus}, {D'Ammando}, {de Palma}, {Digel}, {Di Venere}, {Drell}, {Favuzzi},
  {Ferrara}, {Franckowiak}, {Fusco}, {Gargano}, {Gasparrini}, {Giglietto},
  {Giroletti}, {Glanzman}, {Godfrey}, {Grenier}, {Guiriec}, {Hadasch},
  {Harding}, {Hayashida}, {Hayashi}, {Hays}, {Hewitt}, {Hill}, {Horan}, {Hou},
  {Jogler}, {Johnson}, {Johnson}, {Kerr}, {Kn{\"o}dlseder}, {Kuss}, {Lande},
  {Larsson}, {Latronico}, {Lemoine-Goumard}, {Longo}, {Loparco}, {Lubrano},
  {Malyshev}, {Marelli}, {Mayer}, {Mazziotta}, {Mehault}, {Mizuno}, {Monzani},
  {Morselli}, {Murgia}, {Nemmen}, {Nuss}, {Ohsugi}, {Omodei}, {Orienti},
  {Orlando}, {Paneque}, {Pesce-Rollins}, {Pierbattista}, {Piron}, {Pivato},
  {Porter}, {Rain{\`o}}, {Rando}, {Ray}, {Razzano}, {Reimer}, {Reposeur},
  {Romani}, {Sartori}, {Saz Parkinson}, {Sgr{\`o}}, {Siskind}, {Smith},
  {Spinelli}, {Strong}, {Takahashi}, {Thayer}, {Thompson}, {Tibaldo},
  {Tinivella}, {Torres}, {Tosti}, {Uchiyama}, {Usher}, {Vandenbroucke},
  {Vasileiou}, {Venter}, {Vianello}, {Vitale}, {Winer}, \& {Wood}}]{abb+13}
{Allafort}, A., {Baldini}, L., {Ballet}, J., {et~al.} 2013, \apjl, 777, L2

\bibitem[{Alpar {et~al.}(1982)Alpar, Cheng, Ruderman, \& Shaham}]{acrs82}
Alpar, M.~A., Cheng, A.~F., Ruderman, M.~A., \& Shaham, J. 1982, Nature, 300,
  728

\bibitem[{{Archibald} {et~al.}(2010){Archibald}, {Kaspi}, {Bogdanov},
  {Hessels}, {Stairs}, {Ransom}, \& {McLaughlin}}]{akb+10}
{Archibald}, A.~M., {Kaspi}, V.~M., {Bogdanov}, S., {et~al.} 2010, \apj, 722,
  88

\bibitem[{{Archibald} {et~al.}(2013){Archibald}, {Kaspi}, {Hessels},
  {Stappers}, {Janssen}, \& {Lyne}}]{akh+13}
{Archibald}, A.~M., {Kaspi}, V.~M., {Hessels}, J. W.~T., {et~al.} 2013,
  arXiv:1311.5161, submitted

\bibitem[{{Archibald} {et~al.}(2009){Archibald}, {Stairs}, {Ransom}, {Kaspi},
  {Kondratiev}, {Lorimer}, {McLaughlin}, {Boyles}, {Hessels}, {Lynch}, {van
  Leeuwen}, {Roberts}, {Jenet}, {Champion}, {Rosen}, {Barlow}, {Dunlap}, \&
  {Remillard}}]{asr+09}
{Archibald}, A.~M., {Stairs}, I.~H., {Ransom}, S.~M., {et~al.} 2009, Science,
  324, 1411

\bibitem[{{Atwood} {et~al.}(2009){Atwood}, {Abdo}, {Ackermann}, {Althouse},
  {Anderson}, {Axelsson}, {Baldini}, {Ballet}, {Band}, {Barbiellini}, \&
  et~al.}]{aaa+09c}
{Atwood}, W.~B., {Abdo}, A.~A., {Ackermann}, M., {et~al.} 2009, \apj, 697, 1071

\bibitem[{Backer {et~al.}(1982)Backer, Kulkarni, Heiles, Davis, \&
  Goss}]{bkh+82}
Backer, D.~C., Kulkarni, S.~R., Heiles, C., Davis, M.~M., \& Goss, W.~M. 1982,
  Nature, 300, 615

\bibitem[{{Bailes} {et~al.}(2011){Bailes}, {Bates}, {Bhalerao}, {Bhat},
  {Burgay}, {Burke-Spolaor}, {D'Amico}, {Johnston}, {Keith}, {Kramer},
  {Kulkarni}, {Levin}, {Lyne}, {Milia}, {Possenti}, {Spitler}, {Stappers}, \&
  {van Straten}}]{bbb+11}
{Bailes}, M., {Bates}, S.~D., {Bhalerao}, V., {et~al.} 2011, Science, 333, 1717

\bibitem[{{Bogdanov} {et~al.}(2011){Bogdanov}, {Archibald}, {Hessels}, {Kaspi},
  {Lorimer}, {McLaughlin}, {Ransom}, \& {Stairs}}]{bah+11}
{Bogdanov}, S., {Archibald}, A.~M., {Hessels}, J.~W.~T., {et~al.} 2011, \apj,
  742, 97

\bibitem[{{Bond} {et~al.}(2002){Bond}, {White}, {Becker}, \&
  {O'Brien}}]{bwb+02}
{Bond}, H.~E., {White}, R.~L., {Becker}, R.~H., \& {O'Brien}, M.~S. 2002,
  \pasp, 114, 1359

\bibitem[{{Breton} {et~al.}(2013){Breton}, {van Kerkwijk}, {Roberts},
  {Hessels}, {Camilo}, {McLaughlin}, {Ransom}, {Ray}, \& {Stairs}}]{bvr+13}
{Breton}, R.~P., {van Kerkwijk}, M.~H., {Roberts}, M.~S.~E., {et~al.} 2013,
  \apj, 769, 108

\bibitem[{Burderi {et~al.}(2001)Burderi, Possenti, D'Antona, Di~Salvo, Burgay,
  Stella, Menna, Iaria, Compana, \& D'Amico}]{bpd+01}
Burderi, L., Possenti, A., D'Antona, F., {et~al.} 2001, ApJ, 560, L71

\bibitem[{{Campana} {et~al.}(1998){Campana}, {Stella}, {Mereghetti}, {Colpi},
  {Tavani}, {Ricci}, {Fiume}, \& {Belloni}}]{csm+98}
{Campana}, S., {Stella}, L., {Mereghetti}, S., {et~al.} 1998, ApJ, 499, L65+

\bibitem[{{Campana} {et~al.}(2004){Campana}, {D'Avanzo}, {Casares}, {Covino},
  {Israel}, {Marconi}, {Hynes}, {Charles}, \& {Stella}}]{cdc+04}
{Campana}, S., {D'Avanzo}, P., {Casares}, J., {et~al.} 2004, \apjl, 614, L49

\bibitem[{{Chakrabarty} {et~al.}(2003){Chakrabarty}, {Morgan}, {Muno},
  {Galloway}, {Wijnands}, {van der Klis}, \& {Markwardt}}]{cmm+03}
{Chakrabarty}, D., {Morgan}, E.~H., {Muno}, M.~P., {et~al.} 2003, Nature, 424,
  42

\bibitem[{{Cordes} {et~al.}(2006){Cordes}, {Freire}, {Lorimer}, {Camilo},
  {Champion}, {Nice}, {Ramachandran}, {Hessels}, {Vlemmings}, {van Leeuwen},
  {Ransom}, {Bhat}, {Arzoumanian}, {McLaughlin}, {Kaspi}, {Kasian}, {Deneva},
  {Reid}, {Chatterjee}, {Han}, {Backer}, {Stairs}, {Deshpande}, \&
  {Faucher-Gigu{\`e}re}}]{cfl+06}
{Cordes}, J.~M., {Freire}, P.~C.~C., {Lorimer}, D.~R., {et~al.} 2006, ApJ, 637,
  446

\bibitem[{{Deller} {et~al.}(2012){Deller}, {Archibald}, {Brisken},
  {Chatterjee}, {Janssen}, {Kaspi}, {Lorimer}, {Lyne}, {McLaughlin}, {Ransom},
  {Stairs}, \& {Stappers}}]{dab+12}
{Deller}, A.~T., {Archibald}, A.~M., {Brisken}, W.~F., {et~al.} 2012, \apjl,
  756, L25

\bibitem[{{Dubus}(2013)}]{dub13}
{Dubus}, G. 2013, \aapr, 21, 64

\bibitem[{{Ek{\c s}{\.I}} \& {Alpar}(2005)}]{ea05}
{Ek{\c s}{\.I}}, K.~Y., \& {Alpar}, M.~A. 2005, \apj, 620, 390

\bibitem[{Fruchter {et~al.}(1990)Fruchter, Berman, Bower, Convery, Goss,
  Hankins, Klein, Nice, Ryba, Stinebring, Taylor, Thorsett, \&
  Weisberg}]{fbb+90}
Fruchter, A.~S., Berman, G., Bower, G., {et~al.} 1990, ApJ, 351, 642

\bibitem[{Halpern {et~al.}(2013)Halpern, Gaidos, A., \& Price-Whelan}]{hgs+13}
Halpern, J.~P., Gaidos, E., A., S., \& Price-Whelan, A.~M.and~Bogdanov, S.
  2013, The Astronomer's Telegram, 5514, 1

\bibitem[{{Hermsen} {et~al.}(2013){Hermsen}, {Hessels}, {Kuiper}, {van
  Leeuwen}, {Mitra}, {de Plaa}, {Rankin}, {Stappers}, {Wright}, {Basu},
  {Alexov}, {Coenen}, {Grie{\ss}meier}, {Hassall}, {Karastergiou}, {Keane},
  {Kondratiev}, {Kramer}, {Kuniyoshi}, {Noutsos}, {Serylak}, {Pilia}, {Sobey},
  {Weltevrede}, {Zagkouris}, {Asgekar}, {Avruch}, {Batejat}, {Bell}, {Bell},
  {Bentum}, {Bernardi}, {Best}, {B{\^i}rzan}, {Bonafede}, {Breitling},
  {Broderick}, {Br{\"u}ggen}, {Butcher}, {Ciardi}, {Duscha}, {Eisl{\"o}ffel},
  {Falcke}, {Fender}, {Ferrari}, {Frieswijk}, {Garrett}, {de Gasperin}, {de
  Geus}, {Gunst}, {Heald}, {Hoeft}, {Horneffer}, {Iacobelli}, {Kuper}, {Maat},
  {Macario}, {Markoff}, {McKean}, {Mevius}, {Miller-Jones}, {Morganti}, {Munk},
  {Orr{\'u}}, {Paas}, {Pandey-Pommier}, {Pandey}, {Pizzo}, {Polatidis},
  {Rawlings}, {Reich}, {R{\"o}ttgering}, {Scaife}, {Schoenmakers}, {Shulevski},
  {Sluman}, {Steinmetz}, {Tagger}, {Tang}, {Tasse}, {ter Veen}, {Vermeulen},
  {van de Brink}, {van Weeren}, {Wijers}, {Wise}, {Wucknitz}, {Yatawatta}, \&
  {Zarka}}]{hhk+13}
{Hermsen}, W., {Hessels}, J.~W.~T., {Kuiper}, L., {et~al.} 2013, Science, 339,
  436

\bibitem[{{Hessels} {et~al.}(2008){Hessels}, {Ransom}, {Kaspi}, {Roberts},
  {Champion}, \& {Stappers}}]{hrk+08}
{Hessels}, J.~W.~T., {Ransom}, S.~M., {Kaspi}, V.~M., {et~al.} 2008, in
  American Institute of Physics Conference Series, Vol. 983, 40 Years of
  Pulsars: Millisecond Pulsars, Magnetars and More, ed. C.~{Bassa}, Z.~{Wang},
  A.~{Cumming}, \& V.~M. {Kaspi}, 613--615

\bibitem[{{Hessels} {et~al.}(2006){Hessels}, {Ransom}, {Stairs}, {Freire},
  {Kaspi}, \& {Camilo}}]{hrs+06}
{Hessels}, J.~W.~T., {Ransom}, S.~M., {Stairs}, I.~H., {et~al.} 2006, Science,
  311, 1901

\bibitem[{Hessels {et~al.}(2007)Hessels, Ransom, Stairs, Kaspi, \&
  Freire}]{hrs+07}
Hessels, J. W.~T., Ransom, S.~M., Stairs, I.~H., Kaspi, V.~M., \& Freire, P.
  C.~C. 2007, ApJ, 670, 363

\bibitem[{{Hessels} {et~al.}(2011){Hessels}, {Roberts}, {McLaughlin}, {Ray},
  {Bangale}, {Ransom}, {Kerr}, {Camilo}, \& {Decesar}}]{hrm+11}
{Hessels}, J.~W.~T., {Roberts}, M.~S.~E., {McLaughlin}, M.~A., {et~al.} 2011,
  in American Institute of Physics Conference Series, Vol. 1357, American
  Institute of Physics Conference Series, ed. M.~{Burgay}, N.~{D'Amico},
  P.~{Esposito}, A.~{Pellizzoni}, \& A.~{Possenti}, 40--43

\bibitem[{{Hobbs} {et~al.}(2006){Hobbs}, {Edwards}, \& {Manchester}}]{hem06}
{Hobbs}, G.~B., {Edwards}, R.~T., \& {Manchester}, R.~N. 2006, MNRAS, 369, 655

\bibitem[{{Homer} {et~al.}(2006){Homer}, {Szkody}, {Chen}, {Henden}, {Schmidt},
  {Anderson}, {Silvestri}, \& {Brinkmann}}]{hsc+06}
{Homer}, L., {Szkody}, P., {Chen}, B., {et~al.} 2006, \aj, 131, 562

\bibitem[{{Karuppusamy} {et~al.}(2008){Karuppusamy}, {Stappers}, \& {van
  Straten}}]{ksv08}
{Karuppusamy}, R., {Stappers}, B., \& {van Straten}, W. 2008, \pasp, 120, 191

\bibitem[{{Keith} {et~al.}(2010){Keith}, {Jameson}, {van Straten}, {Bailes},
  {Johnston}, {Kramer}, {Possenti}, {Bates}, {Bhat}, {Burgay}, {Burke-Spolaor},
  {D'Amico}, {Levin}, {McMahon}, {Milia}, \& {Stappers}}]{kjv+10}
{Keith}, M.~J., {Jameson}, A., {van Straten}, W., {et~al.} 2010, \mnras, 1356

\bibitem[{{Kerr}(2011)}]{ker11}
{Kerr}, M. 2011, \apj, 732, 38

\bibitem[{{Kramer} {et~al.}(2006){Kramer}, {Lyne}, {O'Brien}, {Jordan}, \&
  {Lorimer}}]{klo+06}
{Kramer}, M., {Lyne}, A.~G., {O'Brien}, J.~T., {Jordan}, C.~A., \& {Lorimer},
  D.~R. 2006, Science, 312, 549

\bibitem[{{Krimm} {et~al.}(2013){Krimm}, {Holland}, {Corbet}, {Pearlman},
  {Romano}, {Kennea}, {Bloom}, {Barthelmy}, {Baumgartner}, {Cummings},
  {Gehrels}, {Lien}, {Markwardt}, {Palmer}, {Sakamoto}, {Stamatikos}, \&
  {Ukwatta}}]{khc+13}
{Krimm}, H.~A., {Holland}, S.~T., {Corbet}, R.~H.~D., {et~al.} 2013, \apjs,
  209, 14

\bibitem[{{Linares} {et~al.}(2013){Linares}, {Bahramian}, {Heinke}, {Wijnands},
  {Patruno}, {Altamirano}, {Homan}, {Bogdanov}, \& {Pooley}}]{lbh+13}
{Linares}, M., {Bahramian}, A., {Heinke}, C., {et~al.} 2013, ArXiv e-prints,
  arXiv:1310.7937

\bibitem[{{Lyne} {et~al.}(2010){Lyne}, {Hobbs}, {Kramer}, {Stairs}, \&
  {Stappers}}]{lhk+10}
{Lyne}, A., {Hobbs}, G., {Kramer}, M., {Stairs}, I., \& {Stappers}, B. 2010,
  Science, 329, 408

\bibitem[{{Nolan} {et~al.}(2012){Nolan}, {Abdo}, {Ackermann}, {Ajello},
  {Allafort}, {Antolini}, {Atwood}, {Axelsson}, {Baldini}, {Ballet}, \&
  et~al.}]{naa+12}
{Nolan}, P.~L., {Abdo}, A.~A., {Ackermann}, M., {et~al.} 2012, \apjs, 199, 31

\bibitem[{{Papitto} {et~al.}(2013){Papitto}, {Ferrigno}, {Bozzo}, {Rea},
  {Pavan}, {Burderi}, {Burgay}, {Campana}, {di Salvo}, {Falanga},
  {Filipovi{\'c}}, {Freire}, {Hessels}, {Possenti}, {Ransom}, {Riggio},
  {Romano}, {Sarkissian}, {Stairs}, {Stella}, {Torres}, {Wieringa}, \&
  {Wong}}]{pfb+13}
{Papitto}, A., {Ferrigno}, C., {Bozzo}, E., {et~al.} 2013, \nat, 501, 517

\bibitem[{{Patruno} \& {Watts}(2012)}]{pw12}
{Patruno}, A., \& {Watts}, A.~L. 2012, ArXiv e-prints, arXiv:1206.2727

\bibitem[{{Patruno} {et~al.}(2013){Patruno}, {Archibald}, {Hessels},
  {Bogdanov}, {Stappers}, {Bassa}, {Janssen}, {Kaspi}, {Tendulkar}, \&
  {Lyne}}]{pah+13}
{Patruno}, A., {Archibald}, A.~M., {Hessels}, J.~W.~T., {et~al.} 2013, ArXiv
  e-prints, arXiv:1310.7549

\bibitem[{Radhakrishnan \& Srinivasan(1981)}]{rs81}
Radhakrishnan, V., \& Srinivasan, G. 1981, in Proc. 2nd Asian--Pacific Regional
  Meeting of the IAU, ed. B.~Hidayat \& M.~W. Feast (Jakarta: Tira Pustaka),
  423--432

\bibitem[{{Ransom}(2013)}]{ran13}
{Ransom}, S.~M. 2013, in IAU Symposium, Vol. 291, IAU Symposium, 3--10

\bibitem[{{Ray} {et~al.}(2012){Ray}, {Abdo}, {Parent}, {Bhattacharya},
  {Bhattacharyya}, {Camilo}, {Cognard}, {Theureau}, {Ferrara}, {Harding},
  {Thompson}, {Freire}, {Guillemot}, {Gupta}, {Roy}, {Hessels}, {Johnston},
  {Keith}, {Shannon}, {Kerr}, {Michelson}, {Romani}, {Kramer}, {McLaughlin},
  {Ransom}, {Roberts}, {Saz Parkinson}, {Ziegler}, {Smith}, {Stappers},
  {Weltevrede}, \& {Wood}}]{rap+12}
{Ray}, P.~S., {Abdo}, A.~A., {Parent}, D., {et~al.} 2012, ArXiv e-prints,
  arXiv:1205.3089

\bibitem[{{Ray} {et~al.}(2013){Ray}, {Ransom}, {Cheung}, {Giroletti},
  {Cognard}, {Camilo}, {Bhattacharyya}, {Roy}, {Romani}, {Ferrara},
  {Guillemot}, {Johnston}, {Keith}, {Kerr}, {Kramer}, {Pletsch}, {Saz
  Parkinson}, \& {Wood}}]{rrc+13}
{Ray}, P.~S., {Ransom}, S.~M., {Cheung}, C.~C., {et~al.} 2013, \apjl, 763, L13

\bibitem[{{Roberts}(2011)}]{rob11}
{Roberts}, M.~S.~E. 2011, in American Institute of Physics Conference Series,
  Vol. 1357, American Institute of Physics Conference Series, ed. M.~{Burgay},
  N.~{D'Amico}, P.~{Esposito}, A.~{Pellizzoni}, \& A.~{Possenti}, 127--130

\bibitem[{{Romani} \& {Shaw}(2011)}]{rs11}
{Romani}, R.~W., \& {Shaw}, M.~S. 2011, \apjl, 743, L26

\bibitem[{{Ryba} \& {Taylor}(1991)}]{rt91b}
{Ryba}, M.~F., \& {Taylor}, J.~H. 1991, ApJ, 380, 557

\bibitem[{{Shvartsman}(1970)}]{shv70}
{Shvartsman}, V.~F. 1970, \sovast, 14, 527

\bibitem[{Stappers {et~al.}(1996)Stappers, Bailes, Lyne, Manchester, D'Amico,
  Tauris, Lorimer, Johnston, \& Sandhu}]{sbl+96}
Stappers, B.~W., Bailes, M., Lyne, A.~G., {et~al.} 1996, ApJ, 465, L119

\bibitem[{{Tam} {et~al.}(2010){Tam}, {Hui}, {Huang}, {Kong}, {Takata}, {Lin},
  {Yang}, {Cheng}, \& {Taam}}]{thh+10}
{Tam}, P.~H.~T., {Hui}, C.~Y., {Huang}, R.~H.~H., {et~al.} 2010, \apjl, 724,
  L207

\bibitem[{{Tauris}(2012)}]{tau12}
{Tauris}, T.~M. 2012, Science, 335, 561

\bibitem[{Tavani(1991)}]{tav91}
Tavani, M. 1991, Nature, 351, 39

\bibitem[{Tavani(1993)}]{tav93}
---. 1993, ApJ, 407, 135

\bibitem[{{Thorstensen} \& {Armstrong}(2005)}]{ta05}
{Thorstensen}, J.~R., \& {Armstrong}, E. 2005, \aj, 130, 759

\bibitem[{{van Haaften} {et~al.}(2012){van Haaften}, {Nelemans}, {Voss}, \&
  {Jonker}}]{hnvj12}
{van Haaften}, L.~M., {Nelemans}, G., {Voss}, R., \& {Jonker}, P.~G. 2012,
  \aap, 541, A22

\bibitem[{{van Straten} \& {Bailes}(2011)}]{vb10}
{van Straten}, W., \& {Bailes}, M. 2011, PASA, 28, 1

\bibitem[{{van Straten} {et~al.}(2012){van Straten}, {Demorest}, \&
  {Oslowski}}]{vdo12}
{van Straten}, W., {Demorest}, P., \& {Oslowski}, S. 2012, Astronomical
  Research and Technology, 9, 237

\bibitem[{Wijnands \& van~der Klis(1998)}]{wv98}
Wijnands, R., \& van~der Klis, M. 1998, Nature, 394, 344

\bibitem[{{Woudt} {et~al.}(2004){Woudt}, {Warner}, \& {Pretorius}}]{wwp+04}
{Woudt}, P.~A., {Warner}, B., \& {Pretorius}, M.~L. 2004, \mnras, 351, 1015

\bibitem[{{Xing} \& {Wang}(2013)}]{xw13}
{Xing}, Y., \& {Wang}, Z. 2013, \apj, 769, 119

\end{thebibliography}
\end{document}